\documentclass[conference]{IEEEtran}
\IEEEoverridecommandlockouts
\usepackage{cite}
\usepackage{amsmath,amssymb,amsfonts}
\usepackage{algorithmic}
\usepackage{graphicx}
\usepackage{textcomp}
\usepackage{xcolor}

\usepackage{url} %志田追加
\usepackage{comment} %志田追加

\usepackage{bm} % Ozaki追加

\def\BibTeX{{\rm B\kern-.05em{\sc i\kern-.025em b}\kern-.08em
    T\kern-.1667em\lower.7ex\hbox{E}\kern-.125emX}}
\begin{document}

%%%%%%%%%%%%%%%%%%%%
\onecolumn
\;\\ \;\\ 
{\Huge \centering
Privacy-Preserving Synthetic Dataset of Individual Daily Trajectories for City-Scale Mobility Analytics\\
}
\vspace{0.5cm} 
{\Large \centering
Jun'ichi Ozaki\textsuperscript{\dag}, Ryosuke Susuta\textsuperscript{\dag}, Takuhiro Moriyama, and Yohei Shida\\
}
\vspace{0.5cm} 
{\Large 
\copyright 2025 IEEE.  Personal use of this material is permitted.  Permission from IEEE must be obtained for all other uses, in any current or future media, including reprinting/republishing this material for advertising or promotional purposes, creating new collective works, for resale or redistribution to servers or lists, or reuse of any copyrighted component of this work in other works.
}

\twocolumn
\newpage
%%%%%%%%%%%%%%%%%%%%

\title{Privacy-Preserving Synthetic Dataset of Individual Daily Trajectories for City-Scale Mobility Analytics\\
%{\footnotesize \textsuperscript{*}Note: Sub-titles are not captured in Xplore and should not be used}
\thanks{This work was supported in part by GEOTRA Co., Ltd. through a collaborative research agreement.}
}

\author{
% 1) First author — Yokohama City University（equal: † を明示）
\IEEEauthorblockN{ Jun'ichi Ozaki\textsuperscript{\dag}}
\IEEEauthorblockA{\textit{School of Data Science} \\
\textit{Yokohama City University}\\
Yokohama, Japan \\
ozaki.jun.fe@yokohama-cu.ac.jp}
\and
% 2) Second author — GEOTRA（equal: †）
\IEEEauthorblockN{ Ryosuke Susuta\textsuperscript{\dag}}
\IEEEauthorblockA{\textit{GEOTRA Co., Ltd.}\\
Tokyo, Japan \\
susuta@geotra.jp}
\and
% 3) Third author — GEOTRA（脚注なし）
\IEEEauthorblockN{ Takuhiro Moriyama}
\IEEEauthorblockA{\textit{GEOTRA Co., Ltd.}\\
Tokyo, Japan \\
moriyama@geotra.jp}
\and
% 4) Fourth author — University of Tsukuba（corresponding: ‡）
\IEEEauthorblockN{Yohei Shida\textsuperscript{\ddag}}
\IEEEauthorblockA{\textit{Institute of Systems and Information Engineering} \\
\textit{University of Tsukuba}\\
Tsukuba, Japan \\
shida@sk.tsukuba.ac.jp}
}

\maketitle

% --- 著者用の注記を記号で本文下に必ず出す ---
\begingroup
\renewcommand{\thefootnote}{\fnsymbol{footnote}} % 1=*, 2=†, 3=‡, 4=§, ...
\footnotetext[2]{These authors contributed equally to this work.} % † ← 1&2に対応
\footnotetext[3]{Corresponding author: shida@sk.tsukuba.ac.jp}              % ‡ ← 4に対応
\endgroup

\begin{abstract}
%結果のところは加筆するかも（まだ出てないので）
Urban mobility data are indispensable for urban planning, transportation demand forecasting, pandemic modeling, and many other applications; however, individual mobile phone-derived Global Positioning System traces cannot generally be shared with third parties owing to severe re-identification risks. Aggregated records, such as origin–destination (OD) matrices, offer partial insights but fail to capture the key behavioral properties of daily human movement, limiting realistic city-scale analyses.

This study presents a privacy-preserving synthetic mobility dataset that reconstructs daily trajectories from aggregated inputs. The proposed method integrates OD flows with two complementary behavioral constraints: (1) dwell–travel time quantiles that are available only as coarse summary statistics and (2) the universal law for the daily distribution of the number of visited locations. Embedding these elements in a multi-objective optimization framework enables the reproduction of realistic distributions of human mobility while ensuring that no personal identifiers are required.

The proposed framework is validated in two contrasting regions of Japan: (1) the 23 special wards of Tokyo, representing a dense metropolitan environment; and (2) Fukuoka Prefecture, where urban and suburban mobility patterns coexist. The resulting synthetic mobility data reproduce dwell-travel time and visit frequency distributions with high fidelity, while deviations in OD consistency remain within the natural range of daily fluctuations.

The results of this study establish a practical synthesis pathway under real-world constraints, providing governments, urban planners, and industries with scalable access to high-resolution mobility data for reliable analytics without the need for sensitive personal records, and supporting practical deployments in policy and commercial domains.
\end{abstract}


\section{Introduction}
Location data serve as an indispensable foundation for decision-making across diverse domains, including urban transportation, logistics, optimization of commercial activities, disaster prevention, and finance \cite{analytics2016age,Unlockin33:online}. Its economic value is substantial, and it has been estimated that the Global Positioning System (GPS) generated approximately USD 1.4 trillion in cumulative economic benefits for the U.S. private sector between 1984 and 2017 \cite{o2019economic}. Furthermore, market forecasts indicate that revenues in the downstream Global Navigation Satellite System (GNSS) market, which includes devices and services, reached EUR 260 billion in 2023 and are expected to double to EUR 580 billion by 2033 \cite{ThenewEU4:online}. Within this expansive market, mobile phone-derived location data have attracted particular attention in recent years, owing to their broad population coverage and high spatio-temporal resolution. In addition to established applications, such as urban transportation planning and commercial site selection, there are expanding uses in new domains, including credit scoring and insurance underwriting \cite{TheEmerg43:online}. However, high-precision GPS data, which can capture detailed individual trajectories at scale, entail significant re-identification risks. Furthermore, when firms provide such data to third parties, they must operate under strict constraints imposed by national privacy laws.

Many countries classify mobile phone-derived location data as personal or sensitive personal information and require data providers to implement measures that effectively eliminate re-identification risks \cite{Dataprot92:online,Californ99:online,LawsandP91:online,location72:online}. Service providers typically combine spatial coarsening (e.g., aggregation into 1 km grids or administrative units), temporal coarsening (e.g., rounding to time intervals or days), and anonymization techniques (e.g., k-anonymity or geomasking) to de-identify data before providing it to third parties \cite{Pseudony54:online}. However, previous research has shown that as few as four spatio-temporal points can uniquely identify an individual\cite{de2013unique}, demonstrating that anonymization alone cannot sufficiently prevent re-identification risks\cite{houssiau2022difficulty}. As a result, coarse-grained data not only hinder organizational operations and detailed analyses but also fail to provide adequate privacy protection. Consequently, companies and municipalities cannot obtain high-resolution individual-level data, including attributes such as age, sex, and behavioral sequences, which makes it difficult to improve the accuracy of advanced urban planning and demand forecasting.
%% However, previous research has shown that as few as four spatio-temporal points can uniquely identify an individual\cite{de2013unique}, demonstrating that anonymization alone cannot sufficiently prevent re-identification risks\cite{houssiau2022difficulty}. camera ready

As a promising approach to overcoming these constraints, synthetic mobility data, defined as data that simulates statistically consistent virtual individuals and their daily trajectories, has recently gained increasing attention in both academia \cite{bougie2025citysim,wu2024imitate} and industry \cite{ReplicaD33:online}. Synthetic data refers to artificially generated data that preserve the statistical and structural properties of the observed datasets. In the context of human mobility, such data enable the use of detailed trajectories while safeguarding privacy. In domains such as healthcare and finance, such data has already been widely adopted \cite{gonzales2023synthetic,jensen2023synthetic}. The synthetic data market is expected to grow at an annual average rate of approximately 30\% until 2030 \cite{Syntheti13:online,Syntheti75:online} and is increasingly regarded as a foundation for enabling detailed analysis and planning under strict privacy constraints. In the human mobility domain, a representative early industrial example is ‘Replica in the United States’, where synthetic mobility data has been widely utilized by U.S. municipalities for transportation policies and commercial site strategies \cite{ReplicaD33:online}. However, many existing approaches depend heavily on specific database schemas or use cases, making it difficult to guarantee performance outside these contexts.

GEOTRA Co., Ltd. provides synthetic mobility data that include one-day mobility trajectories and activity information annotated with sex and age attributes \cite{GEOT42:online}. The data have been adopted by municipalities (including the Tokyo Metropolitan Government) and companies in diverse industries, such as construction, electric power, finance, transportation, and manufacturing, and have been used in urban policy-making and business strategy. The company was established as a joint venture between the KDDI Corporation and Mitsui \& Co., Ltd. \cite{KDDICORP85:online,MITSUICO55:online}. Under strict privacy policies mandated by parent companies, although the final outputs are synthetic, the inputs are restricted to origin-destination (OD) matrices aggregated by sex and age from mobile phone-derived GPS-based mobility data collected by KDDI. Consequently, the consistency criteria used in the generation process are skewed toward the OD metrics, making it difficult to thoroughly validate multiple indicators that capture human-like behavioral properties.

This study proposes an extended approach developed as part of a joint research project with two universities to generate synthetic mobility data that simultaneously align multiple behavioral properties under limited input environments. The approach reconstructs the underlying distributional shapes using quantile statistics of the dwell-travel time distribution obtained from the parent company and further refers to the universal law of the daily distribution of the number of visited locations, as reported in the academic literature. These two distributions are incorporated as auxiliary evaluation metrics into a multi-objective optimization framework, thereby enabling the generation of synthetic mobility data under strict privacy constraints. The case presented in this study demonstrates a practical deployment in which large-scale mobile phone-derived location data owned by a parent company are utilized under strict privacy constraints to produce synthetic mobility data that simultaneously align multiple behavioral properties. These results demonstrate broader applicability in social infrastructure domains that require accurate human mobility forecasts, such as urban transportation planning and commercial site strategies, and provide an early industry use case that simultaneously satisfies reproducibility, generalizability, and privacy protection in the rapidly growing synthetic data market, thereby accelerating applications and research across the industry.

%TODO This section could be improved by adding a short paragraph that details the structure of the paper i.e. ‘The remainder of the paper is organized as follows. In Section 2…’

The remainder of this paper is organized as follows. Section II describes the datasets used in this study, including aggregated OD matrices, the conventional GEOTRA Activity Data, and newly introduced quantile statistics of dwell–travel times. Section III details the proposed multi-objective optimization framework for generating synthetic mobility data. Section IV presents the data generation results obtained for Tokyo and Fukuoka, and Section V discusses the implications, limitations, and opportunities for future work, concluding the paper.

\section{Data}
This section describes the data used to extend the generation process of the synthetic mobility data provided by GEOTRA (GEOTRA Activity Data; GAD). The inputs included aggregated OD matrices derived from mobile phone-derived GPS data, the existing GAD dataset and its generation method, and newly introduced quantile statistics based on the dwell-travel time distribution. The two study areas were: (1) the 23 special wards of Tokyo and (2) Fukuoka Prefecture. Tokyo is the largest metropolitan region in Japan, characterized by a high density and diverse mobility and activity patterns. Fukuoka Prefecture includes a regional core city and surrounding areas where different mobility characteristics emerge from the coexistence of urban and suburban environments. By combining these two regions, this study examined the applicability of the proposed method under varying population densities and urban structures.

\subsection{Aggregated OD Matrices from KDDI}
The aggregated OD matrices used in this study were created by the KDDI Corporation, a major Japanese telecommunications operator, based on mobile phone-derived GPS location data formally provided as the ``KDDI Location Data.'' These data were collected with the explicit consent of subscribers through the mobile phone service brand of KDDI \emph{au}. The collected data consisted of a latitude-longitude time-series recorded every few minutes for each subscriber, with sufficient resolution to reconstruct detailed trajectories. Although the number of GPS data subjects was not disclosed owing to the internal policy of KDDI, the underlying population covers one of the largest subscriber bases in the Japanese mobile phone market, enabling nationwide population coverage. The two study areas were: (1) the 23 special wards of Tokyo and (2) Fukuoka Prefecture. The observation periods were all weekdays in December 2023 for Tokyo and all weekdays in June 2023 for Fukuoka Prefecture. Aggregating data for multiple days smoothed daily fluctuations and ensured statistical stability. A dwell is defined as remaining at the same location for at least 15 min. The native spatial resolution was at the meter level of the latitude-longitude coordinates.

All the high-resolution GPS trajectories were converted into OD matrices within KDDI. In this conversion, continuous movement sequences were segmented into 1 h intervals, and OD pairs were extracted for each interval. Individual IDs were completely removed, and continuity across intervals was not retained, making it impossible to reconstruct a full-day trajectory for any individual.

The OD matrices provided to GEOTRA aggregated, for each hour, the number of movements from origins to destinations (minimum spatial unit: 125 m grid defined by the Japanese Industrial Standards (JIS)\cite{NewEstab59:online}), stratified by sex (male or female) and age group (20s, 30s, 40s, 50s, and 60+). For privacy protection and ethical considerations regarding minors, individuals under 20 years of age were excluded, and cells with movement counts below a predefined threshold were suppressed. Thus, each record indicated the movement of demographic groups between cells and the number of people that moved, in a spatially and temporally coarsened format. Consequently, continuous individual trajectories and identifiers were not included, and the re-identification risk was substantially reduced by this design. The OD matrices provided by KDDI were anonymized on a fine spatial grid, roughly corresponding to the level of residential distributions that are already published in census-based population statistics; therefore, no additional privacy risk regarding home locations arises in this study. %%The OD matrices provided 以下自宅位置対応camera ready

\subsection{Synthetic Mobility Data (GAD)}
As described above, OD matrices represent aggregated mobility counts by demographic attributes for each time period and therefore exclude individual IDs and continuous trajectories. The GAD are synthetic mobility data that reconstruct full-day trajectories and dwell histories of virtual individuals under the constraints imposed by these OD matrices. Each virtual agent was annotated with attributes such as sex, age, and residence, and was assigned detailed information, including destinations, travel modes, purposes, and travel times, in chronological order.

\begin{figure}[!t]
\centering
\includegraphics[width=\columnwidth]{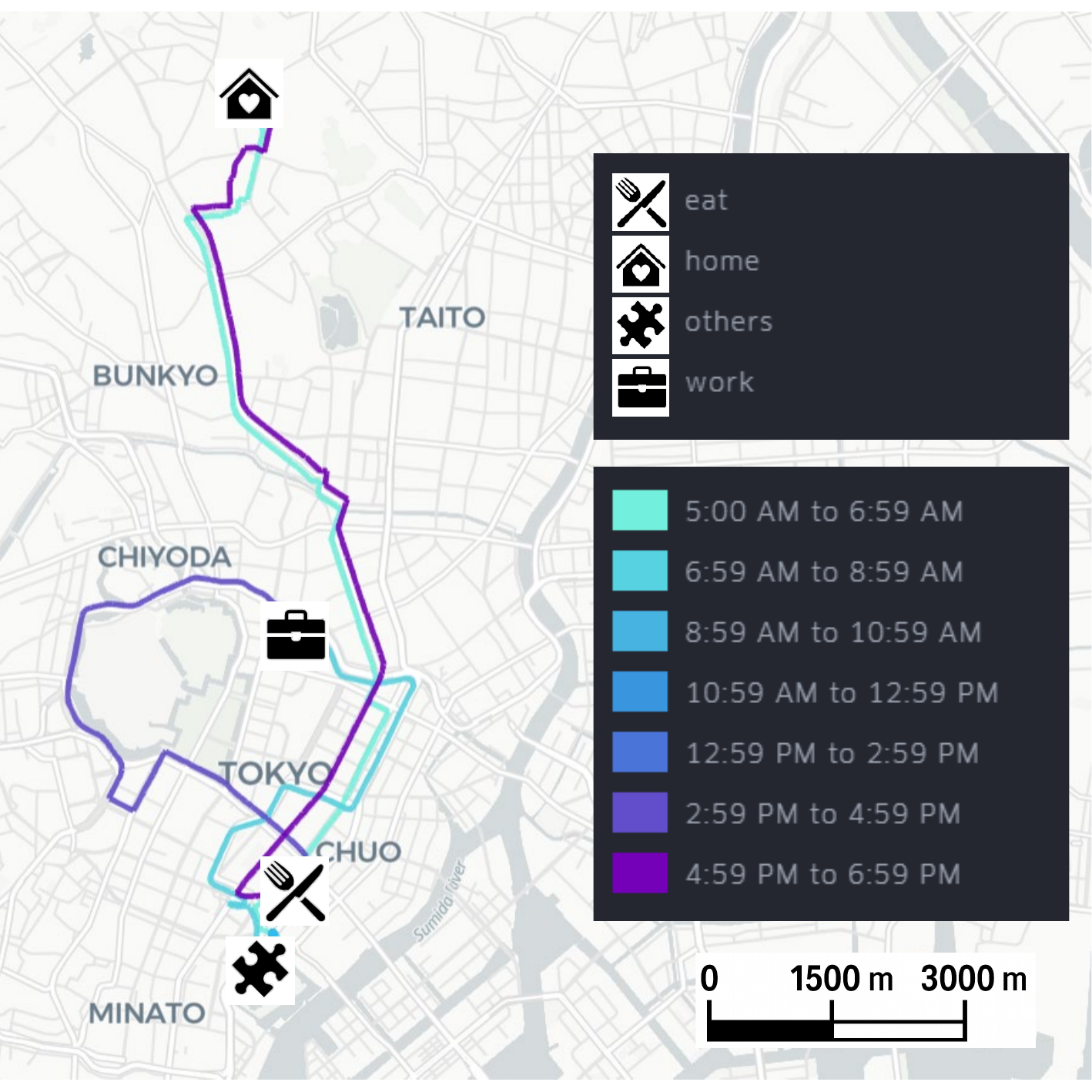}
\caption{\textbf{Example of a daily trajectory from the GAD.}
A virtual user record with assigned age and sex attributes. All movements were by car, with routes allocated using MATSim software on actual road networks. Pictograms denote activity purposes (home, work, eating, and others), and travel segments are color-coded according to the time of day.}
\label{fig:SampleTrajectory}
\end{figure}

Fig. \ref{fig:SampleTrajectory} shows an example of a synthetic daily trajectory included in the GAD. The virtual user was annotated with age and sex attributes and assigned a travel mode and route on actual road networks. In this illustrative case, all trips were by car, and the routes were computed using MATSim software. Activity locations are represented by pictograms (home, work, eating, and others), and travel segments are color-coded according to the time of day.

The generation of the GAD consisted of three stages:
\begin{enumerate}
    \item \textbf{Initialization of the population distribution}: Virtual agents were created in a number equivalent to the total population to match the census-based distributions (sex, age, and residence), and placed at their home locations in the early morning.
    \item \textbf{Temporal evolution and optimization}: Simulated annealing (SA) was applied to satisfy the demographic OD matrices at each time period and determine the mobility trajectory of each agent. Destinations at each step are assigned based on the previous position and OD constraints and were iteratively updated until the entire day was consistent with the OD matrices.
    \item \textbf{Assignment of route information}: For the finalized daily trajectories, MATSim software ~\cite{w2016multi} was used to assign detailed routes on road and public transport networks, including timetables.
\end{enumerate}

Thus, the GAD generated daily virtual individual trajectories by evolving agents over time while ensuring consistency with the OD matrices throughout the day. While conventional GAD emphasizes OD consistency, this process was extended by incorporating additional statistics to reproduce multiple behavioral properties.

\subsection{Quantile Statistics of Dwell-Travel Time}
The quantile statistics of the dwell-travel times are newly provided summary measures obtained through the joint research in this study. Whereas previous frameworks used OD matrices as the sole input, these statistics enabled the incorporation of temporal properties that treat dwell and travel times as a single measure.

These statistics were derived from preprocessed mobile phone-derived GPS data within KDDI. The dwell-travel time is defined as the elapsed duration from arrival at one location to arrival at the next location. In practice, each 125 m grid cell recorded the dwell-travel time as the total duration from the moment a user arrived in that cell to the moment they arrived in the following cell. This interval included both the time spent within the current cell and the time spent traveling to the next destination. Thus, the statistics captured the complete sequence of staying and moving rather than considering them as independent components. For each cell and hour, the distribution was summarized using five quantiles (10\%, 30\%, 50\%, 70\%, and 90\%). The spatial resolution was 125 m and the temporal resolution was in minutes.

The two study areas were: (1) the 23 special wards of Tokyo and (2) Fukuoka Prefecture. In both cases, the statistics were aggregated for all weekdays over one year in 2023. By adopting a long observation period, daily fluctuations were smoothed and statistical stability was ensured. To further preserve privacy and ensure statistical robustness, no demographic stratification (e.g., by age or sex) was applied in the quantile statistics. All users were aggregated and cells with sample sizes below a predefined threshold were suppressed for confidentiality.

\section{Method}

\subsection{Definition and Requirements for Synthetic Data}
This section defines the synthetic mobility data used in this study and clarifies the design requirements, focusing on accurately replicating real-world mobility patterns.  
This study deliberately disregarded the day-by-day fluctuation-level details and instead focused on reproducing the key statistical mean values that were most critical for human mobility analysis.  
Based on this principle, synthetic mobility data should (1) reproduce key expectation values from real data, and (2) match essential distributions or empirical laws of human mobility.
In this study, synthetic mobility data were designed to satisfy these requirements. Specifically, three primary statistics were used: (1) OD matrices, (2) visit frequency distribution, and (3) dwell–travel time distributions. Unlike conventional GAD, which focus solely on OD consistency, the proposed approach integrates the latter two distributions to enhance behavioral realism.

Thus, this study generated full-day trajectories for all virtual individuals in the two target regions: (1) the 23 special wards of Tokyo and (2) Fukuoka Prefecture, as an enhancement of conventional GAD.  All other conditions remained unchanged, except for the explicit incorporation of the visit frequency and dwell–travel time distributions. Using the same spatial and temporal resolution, the synthetic data were refined to more accurately reproduce these two distributions by redesigning Stage 2, ``Temporal evolution and optimization,” as described in the \textit{DATA} section. 

\subsection{Synthetic Data Generation as an Optimization Problem}
The generation of synthetic mobility data was formulated as an optimization problem.  
A loss function is defined to measure the distance between the synthetic and real-world mobility data with respect to key human mobility statistics.  
As described above, this loss function must incorporate discrepancies into three major components: (1) OD matrices, (2) visit frequency distribution, and (3) dwell–travel time distributions.  
Specifically, the following loss terms are defined:
\begin{itemize}
\item $L_\mathrm{OD}$: OD matrix loss
\item $L_\mathrm{VF}$: visit frequency loss
\item $L_\mathrm{DT}$: dwell–travel time loss
\end{itemize}
The total loss is expressed as follows:
\[
L_\mathrm{tot} = w_\mathrm{OD} \frac{L_\mathrm{OD}}{L_\mathrm{OD}(0)} 
+ w_\mathrm{VF} \frac{L_\mathrm{VF}}{L_\mathrm{VF}(0)} 
+ w_\mathrm{DT} \frac{L_\mathrm{DT}}{L_\mathrm{DT}(0)},
\]
where $w_\mathrm{OD}$, $w_\mathrm{VF}$, and $w_\mathrm{DT}$ are the weights assigned to each component, and $L_\mathrm{OD}(0)$, $L_\mathrm{VF}(0)$, and $L_\mathrm{DT}(0)$ denote the normalization factors equal to the initial loss values before the optimization.  
Following the conventional GAD procedure, all virtual agents were initialized at their home locations at $t = 0$ and iteratively optimized until convergence.  
In the baseline GAD, $w_\mathrm{VF} = w_\mathrm{DT} = 0$, meaning that only OD consistency was enforced.  
Increasing these weights enabled the synthetic data to align with additional behavioral constraints.  
Due to data noise and inconsistencies across multiple sources, the loss cannot be minimized to zero, and the weights represent the relative importance of reproducing each statistic.

Although the OD loss function $L_\mathrm{OD}$ in the conventional GAD framework was defined using either the mean squared error (MSE) or the mean absolute error (MAE), depending on the application, an MSE-based formulation was adopted in this study.  
Let $F_{ijt}^{(s)}$ and $F_{ijt}^{(r)}$ denote the OD matrix entries representing the number of flows from origin cell $i$ to destination cell $j$ at time $t$ in the synthetic data (s) and real data (r), respectively.  
The OD loss function is then defined as follows:
\[ L_\mathrm{OD} = \left. \sum_{ijt} c_{ijt} \left(F_{ijt}^{(s)} - F_{ijt}^{(r)}\right)^2 \middle/ \sum_{ijt} \left(F_{ijt}^{(r)}\right)^2 \right., \]
%\[ L_\mathrm{OD} = \frac{\sum_{ijt} c_{ijt} \left(F_{ijt}^{(s)} - F_{ijt}^{(r)}\right)^2}{\sum_{ijt} \left(F_{ijt}^{(r)}\right)^2}, \]
where the summation was performed over all index triplets $(i,j,t)$, and $c_{ijt}$ denotes the error weights:
\[
c_{ijt} =
\begin{cases}
1, & F_{ijt}^{(r)} > 0,\\
15, & F_{ijt}^{(r)} = 0.
\end{cases}
\]
A weight of 15 penalized synthetic OD flows assigned to cell pairs for which no flow was observed in the real data, preventing divergence toward unrealistic connections.  
This formulation can be interpreted as the weighted mean of the squared relative errors between the synthetic and real OD matrices.  
For interpretability, a modified loss is also defined as follows:
\[ L^\mathrm{eval}_\mathrm{OD} = \left. \sum_{ijt} \left(F_{ijt}^{(s)} - F_{ijt}^{(r)}\right)^2 \middle/ \sum_{ijt} \left(F_{ijt}^{(r)}\right)^2 \right., \]
%\[ L^\mathrm{eval}_\mathrm{OD} = \frac{\sum_{ijt} \left(F_{ijt}^{(s)} - F_{ijt}^{(r)}\right)^2}{\sum_{ijt} \left(F_{ijt}^{(r)}\right)^2}, \]
where the square root $\sqrt{L^\mathrm{eval}_\mathrm{OD}}$ denotes the average relative error of the synthetic OD flows.  
$L_\mathrm{OD}$ was used during optimization to penalize unrealistic flows, whereas $L^\mathrm{eval}_\mathrm{OD}$ provided a clearer measure of reporting performance after optimization.

The visit frequency loss function $L_\mathrm{VF}$ was derived from the universal empirical law of human mobility. Schneider \textit{et al.}~\cite{schneider2013unravelling} reported that the daily number of visited locations $N$ per person follows a discrete log-normal distribution:
\[ P^{(r)}(N) \propto N^{-1}e^{-\frac{(\ln N-\mu)^2}{2\sigma^2}}, \quad N \geq 1, \]
%\[ P^{(r)}(N) \propto \frac{1}{N}\exp\left(-\frac{(\ln N-\mu)^2}{2\sigma^2}\right), \quad N \geq 1, \]
with the parameters $\mu = 1$ and $\sigma = 0.5$. The minimum value $N = 1$ occurred only when an individual remained at a single location for 24 h. Using this universal distribution, the visit frequency loss $L_\mathrm{VF}$ is defined as the one-dimensional discrete Wasserstein distance between the synthetic data and empirical law:
\[
L_\mathrm{VF} = D_W(P^{(s)}, P^{(r)}),
\]
where $P^{(s)}(N)$ denotes the probability mass function obtained from the histogram of daily visit counts $N$ in the synthetic data. The Wasserstein distance guaranteed that the difference in the expectations between $P^{(s)}$ and $P^{(r)}$ did not exceed $D_W(P^{(s)}, P^{(r)})$. Thus, $L_\mathrm{VF}$ quantified the maximum error in reproducing the visit frequency distribution in the synthetic data.

For the dwell–travel time loss function $L_\mathrm{DT}$, the dwell–travel time distributions were first parameterized for all the grid cells.  
The dwell–travel time for cell $i$ is defined as the elapsed time from arrival at cell $i$ to arrival at the subsequent cell $j$.  
Equivalently, this metric represents the sum of the dwell time spent in cell $i$ and the travel time from $i$ to the next destination $j$.  
In this study, dwell–travel time statistics were conditioned on both the arrival hour $t$ (ranging from 0 to 23) and the cell index $i$.  
The empirical distribution was summarized using five quantile values: the 10th, 30th, 50th, 70th, and 90th percentiles for each $(i, t)$ pair.  
Only the latter three quantiles (50th, 70th, and 90th) were used to fit the distribution parameters, because the lower quantiles (10th and 30th) are unreliable.  
This unreliability arises from preprocessing of the raw GPS data, which excludes short-term travel segments of less than 15 min (see the \textit{DATA} section).

It was assumed that the dwell–travel time followed a log-normal distribution. 
This choice was motivated by two considerations.
First, Alessandretti \textit{et al.}~\cite{alessandretti2017multi} reported that dwell times follow a log-normal distribution.
Second, typical travel times within both study areas were less than one hour and were typically shorter than the dwell times. 
Consequently, the dwell–travel time distribution, particularly its tail, can be approximated using a log-normal distribution.
The logarithm of the dwell–travel time at the 50th, 70th, and 90th percentiles was approximated as:
\[
\ln T_{i,t}(z) \simeq \mu_{i,t} + \sigma_{i,t} \cdot \mathrm{Probit}(z),
\]
where $\mathrm{Probit}(z)$ is the probit function (i.e., the inverse of the cumulative distribution function of the standard normal distribution) and $z$ corresponds to the quantile levels $z = 0.5, 0.7,$ and $0.9$.  
In addition, $T_{i,t}(z)$ represents the dwell–travel time and $(\mu_{i,t}, \sigma_{i,t})$ are the log-normal parameters defined for each pair of grid cells $i$ and arrival time $t \in \{0,1,\dots,23\}$.  
These parameters were estimated using the least squares method.  
If data were missing for a specific cell $i$, parameter values were assigned based on the neighboring cells using a hierarchical grid approach.  
The 125 $\times$ 125 m grid code defined by the Japanese Industrial Standards (JIS) contains 11 digits, and masking the last digit produces a 250 $\times$ 250 m grid code.  
By iteratively masking the digits, progressively larger cells were obtained, to which the smaller cell belonged.  
For each larger cell, $(\mu_{i,t}, \sigma_{i,t})$ was computed as the mean of the constituent smaller cells.  
If a parameter set for a missing data cell could not be determined directly, it was assigned the parameters of the smallest available cell that contained it and had a defined parameter set.

The dwell–travel time loss function is defined as follows:  
Let $T_{i,t}^{(k)}$ denote the dwell–travel time for agent $k$ at cell $i$ with arrival time $t$ and let $Q_{i,t}(T)$ represent the log-normal distribution at $(i, t)$ estimated from real data.  
As travel durations shorter than 15 min were excluded during preprocessing, the support for dwell–travel times was restricted to $T \geq T_\mathrm{min} = 15\;\mathrm{minutes}$.  
The truncated distribution $\tilde{Q}_{i,t}(T)$ is then as follows:
\[
\tilde{Q}_{i,t}(T) \propto Q_{i,t}(T) \; (T \geq T_\mathrm{min}), \quad 
\tilde{Q}_{i,t}(T) = 0 \; (T < T_\mathrm{min}).
\]
Using this truncated distribution, the Wasserstein distance between the synthetic and empirical dwell–travel time distributions at $(i,t)$ is given by:
\begin{eqnarray*}
D_W(\{T_{i,t}^{(k)}\}_{k},\tilde{Q}_{i,t}) = \qquad\qquad\qquad\qquad\qquad\qquad\qquad\quad \\
\frac{1}{n_{i,t}}\sum_i \left| T_{i,t}^{(k)} - F^{-1}\left( (1-F(T_\mathrm{min}))\frac{i+0.5}{n_{i,t}} + F(T_\mathrm{min}) \right)\right|, 
\end{eqnarray*}
where $n_{i,t}$ is the number of agents at $(i,t)$ and $F$ is the cumulative distribution function of the full log-normal distribution $Q_{i,t}(T)$.  
Finally, the overall dwell–travel time loss is defined as follows:
%\[ L_\mathrm{DT} = \frac{\sum_{i,t} n_{i,t} \, D_W\big(\{T_{i,t}^{(k)}\}_{k}, \tilde{Q}_{i,t}\big)}{\sum_{i,t} n_{i,t}}, \]
\[ L_\mathrm{DT} = \left. {\sum_{i,t} n_{i,t} \, D_W\big(\{T_{i,t}^{(k)}\}_{k}, \tilde{Q}_{i,t}\big)} \middle/ {\sum_{i,t} n_{i,t}} \right. , \]
which represent the weighted mean of Wasserstein distances, where the weights were proportional to the number of agents in each $(i,t)$.  
This metric captured the typical error expected when sampling dwell–travel times from synthetic data instead of real data.

\subsection{Optimization Method and Conditions}
The SA method, as used in the conventional GAD framework, was employed to optimize the loss function defined above and generate synthetic data.  
Following the procedure outlined in the \textit{DATA} section, all virtual agents were initialized at their home locations at $\tau = 0$, with no movement during the day.  
The optimization processed then proceeded through a virtual temporal evolution over discrete simulation steps $\tau$, during which the agent trajectories were iteratively modified.  
The annealing schedule was designed as follows. The temperature was initialized to its maximum at $\tau = 0$, gradually decreased until $\tau = \tau_\mathrm{max}/2$, and reset once to repeat the process until $\tau = \tau_\mathrm{max}$.  
After completing the SA optimization, detailed route information were assigned using the same procedure as in the conventional GAD framework.  
Data generation was performed independently for each attribute (sex and age group) and was subsequently aggregated.

Multi-objective optimization was performed over the three loss functions $(L_\mathrm{OD}, L_\mathrm{VF}, L_\mathrm{DT})$.  
To investigate the sensitivity of the results to the weight parameters $(w_\mathrm{OD}, w_\mathrm{VF}, w_\mathrm{DT})$, a grid search was conducted on a logarithmic scale for each parameter.  
The parameter ranges are defined as follows:  
\begin{itemize}
\item $w_\mathrm{OD} = 1$
\item $w_\mathrm{VF} \in \{0, 0.001, 0.01, 0.1\}$
\item $w_\mathrm{DT} \in \{0, 0.02, 0.05, 0.1, 0.2, 0.5, 1, 2\}$
\end{itemize}
The parameter settings $(w_\mathrm{OD}, w_\mathrm{VF}, w_\mathrm{DT}) = (1, 0, 0)$ correspond to the baseline case used in the conventional GAD framework.  
Increasing $w_\mathrm{VF}$ and $w_\mathrm{DT}$ is expected to yield synthetic data that more accurately reproduce the visit frequency and dwell–travel time distributions.

\section{Results}

\subsection{Optimization Procedure}

\begin{figure}[!t]
\centering
\includegraphics[width=\columnwidth]{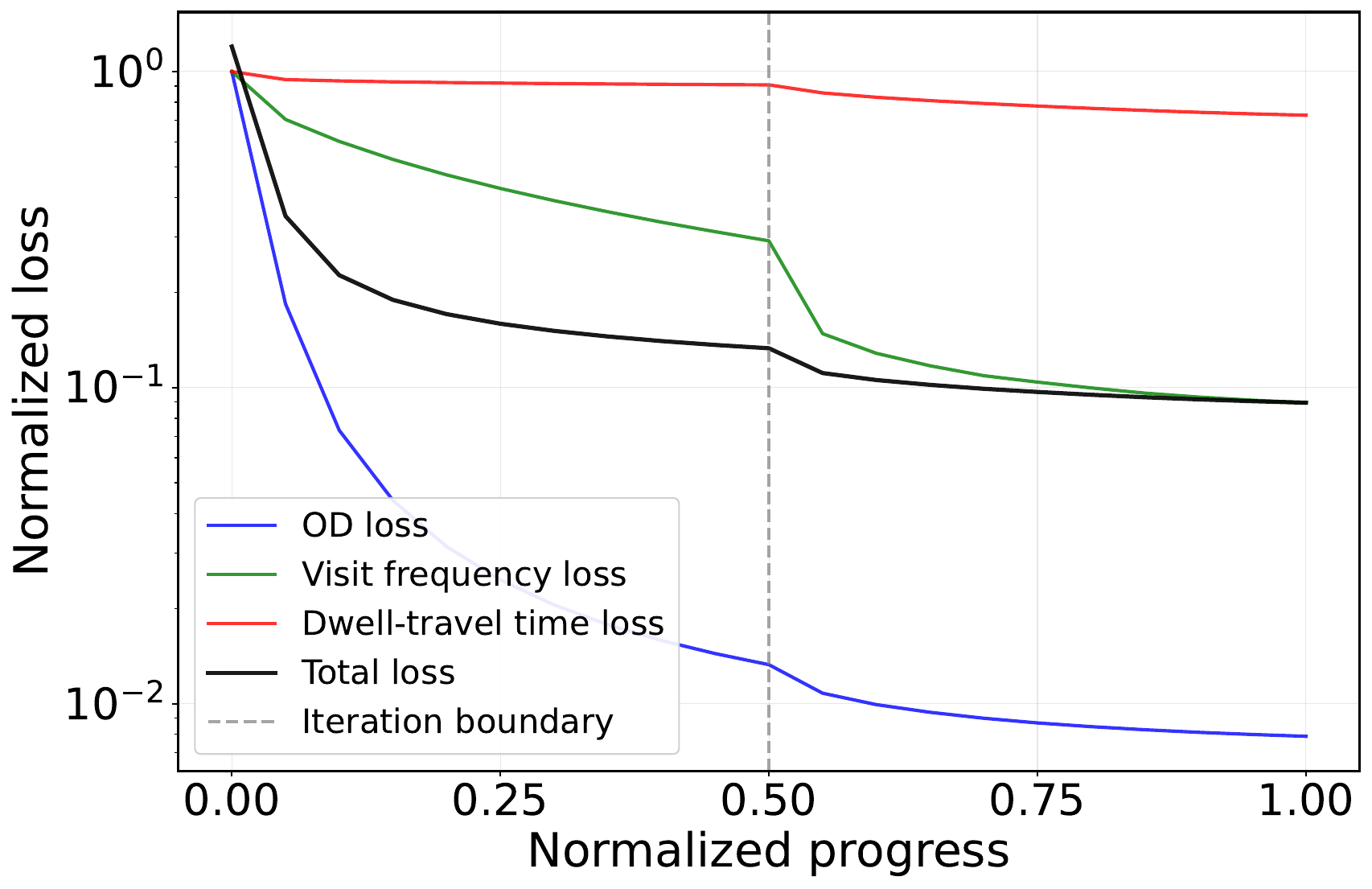}
\caption{\textbf{Optimization procedure.}  
Normalized loss functions $(L_\mathrm{OD}/L_\mathrm{OD}(0), L_\mathrm{VF}/L_\mathrm{VF}(0), L_\mathrm{DT}/L_\mathrm{DT}(0), L_\mathrm{tot})$ are shown for male agents in their twenties, simulated at the parameter settings $(w_\mathrm{OD}, w_\mathrm{VF}, w_\mathrm{DT}) = (1, 0.01, 0.02)$.  
The simulation step $\tau$ was normalized to $[0,1]$. The SA process began at the maximum temperature ($\tau = 0$), decreased until $\tau = 0.5$, and repeated this schedule once, ending at $\tau = 1$. The iteration boundary at $\tau = 0.5$ marked the reset of the temperature to its initial value.}
\label{fig:TotalLossAndTimeStep}
\end{figure}

The optimization process is illustrated in Fig. ~\ref{fig:TotalLossAndTimeStep}, which plots the normalized loss functions (total, OD, visit frequency, and dwell–travel time) for male agents in their twenties at the parameter settings $(w_\mathrm{OD}, w_\mathrm{VF}, w_\mathrm{DT}) = (1, 0.01, 0.02)$.  
The loss values decreased steadily over the SA steps, with a noticeable drop at the midpoint, owing to the scheduled temperature reset.
%20251104 ozaki 追記
The computation time for the optimization process is approximately 20 hours when using a single CPU per attribute (sex and age group) for both Tokyo and Fukuoka. 
%Hence, the total computational cost amounts to about 200 CPU hours for each study area. Regarding computational complexity, the runtime is expected to scale proportionally with the number of agents, as the average number of annealing steps required for convergence does not strongly depend on the agent population.

\subsection{Multi-Objective Optimization}

\begin{figure*}[!t]
\centering
(a)
\includegraphics[width=0.30\linewidth,bb=0 0 1100 900]{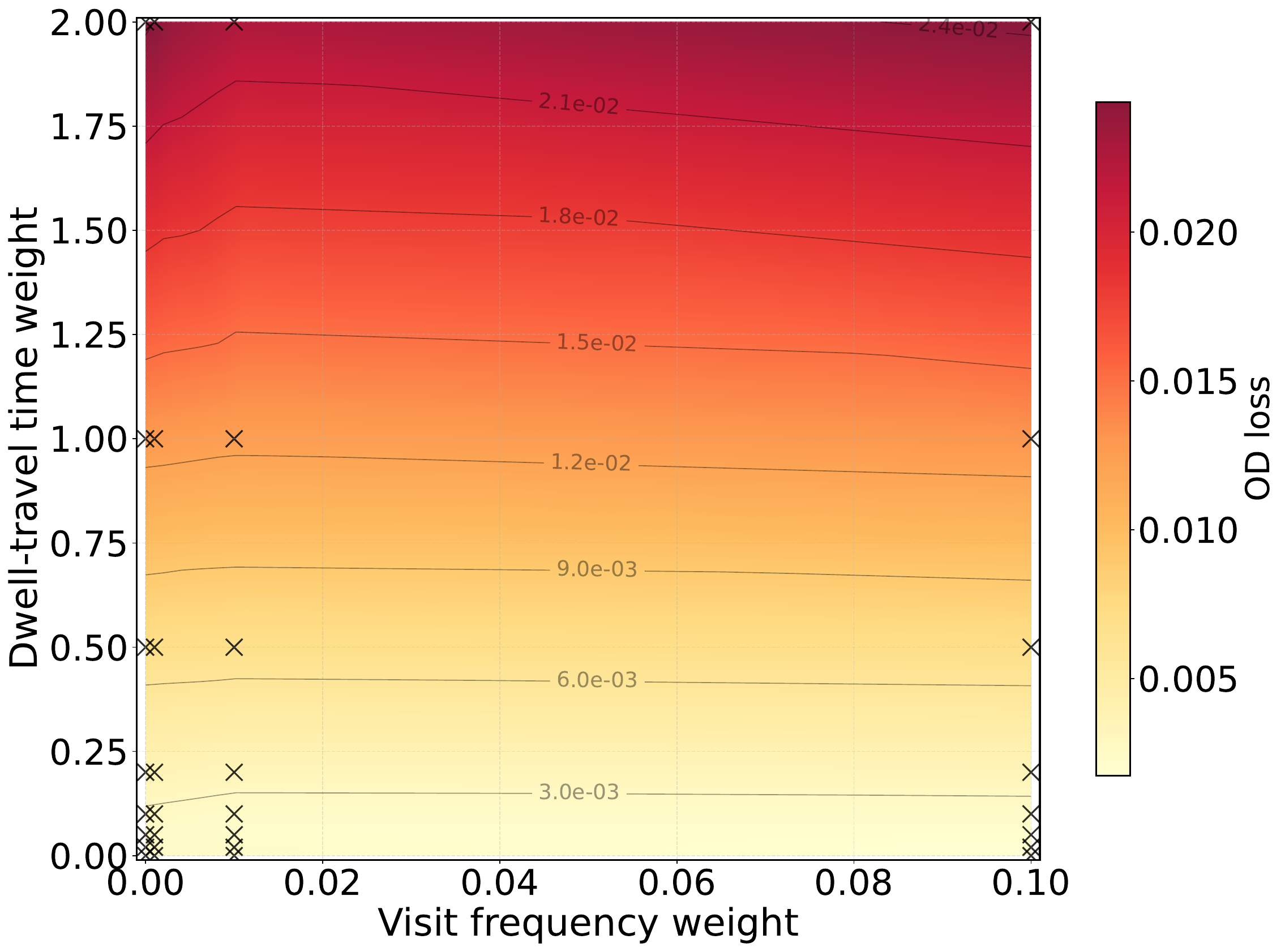}
(b)
\includegraphics[width=0.30\linewidth,bb=0 0 1100 900]{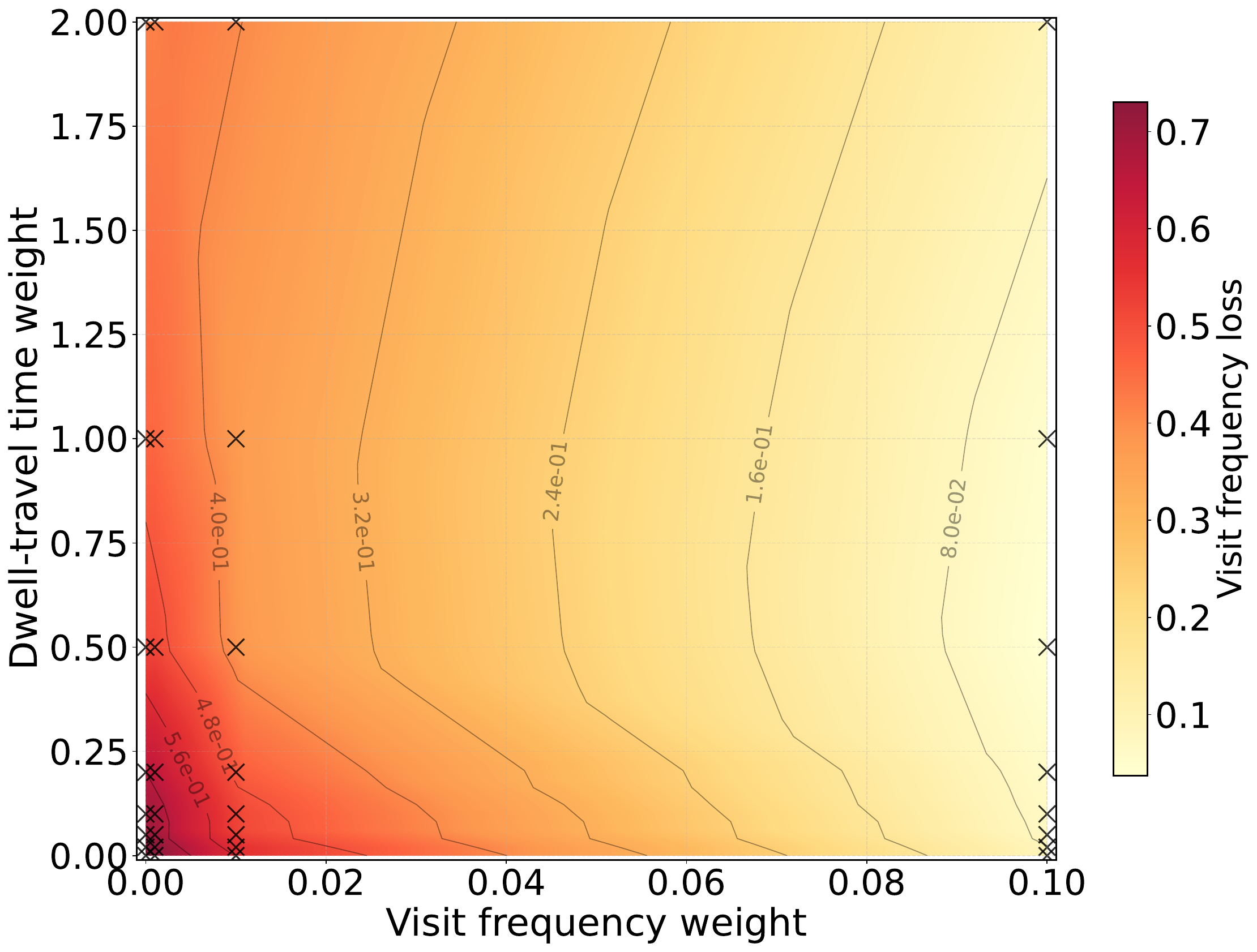}
(c)
\includegraphics[width=0.30\linewidth,bb=0 0 1100 900]{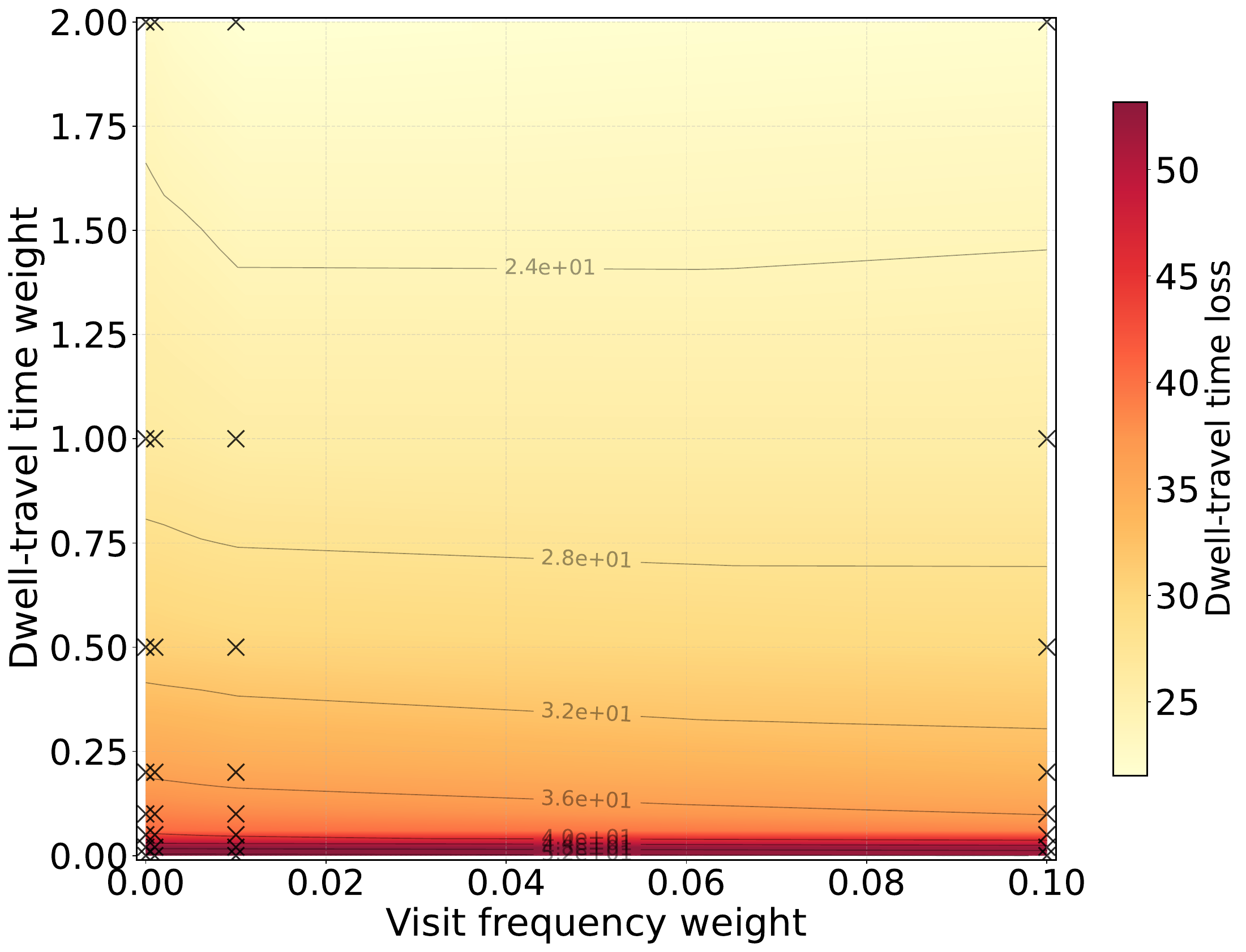}
\caption{\textbf{Loss functions $L^\mathrm{eval}_\mathrm{OD}$, $L_\mathrm{VF}$, and $L_\mathrm{DT}$ after optimization for the 23 special wards of Tokyo.}  
Each loss function was averaged over all attributes (sex and age groups). 
The horizontal and vertical axes represent $w_\mathrm{VF}$ and $w_\mathrm{DT}$, respectively. 
Black crosses indicate the simulated parameter combinations within the grid search range. All three plots demonstrated that the corresponding loss decreased as its associated weight increased.}
\label{fig:LossFunctionsTokyo}
\end{figure*}

\begin{figure*}[!t]
\centering
(a)
\includegraphics[width=0.30\linewidth,bb=0 0 1100 900]{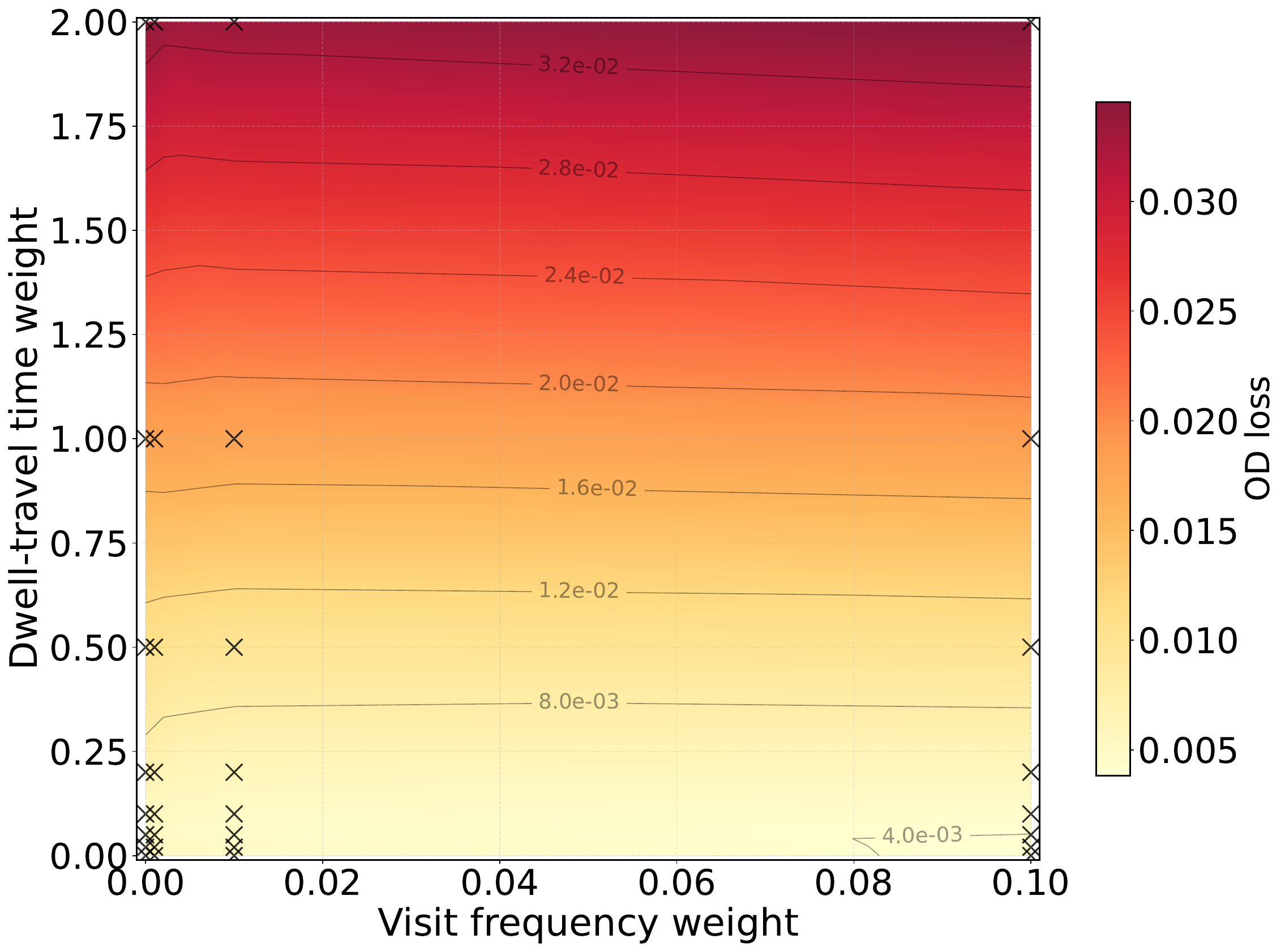}
(b)
\includegraphics[width=0.30\linewidth,bb=0 0 1100 900]{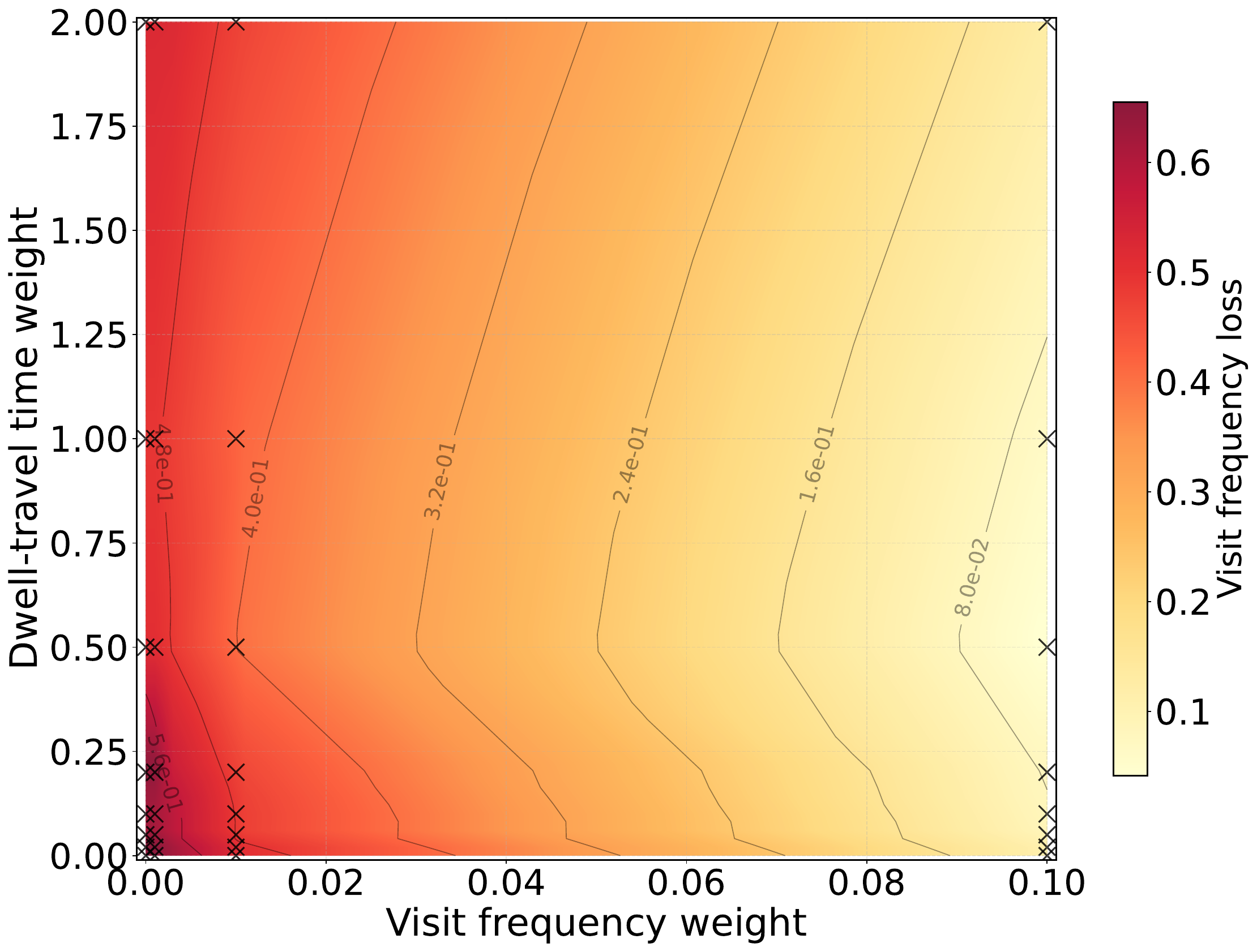}
(c)
\includegraphics[width=0.30\linewidth,bb=0 0 1100 900]{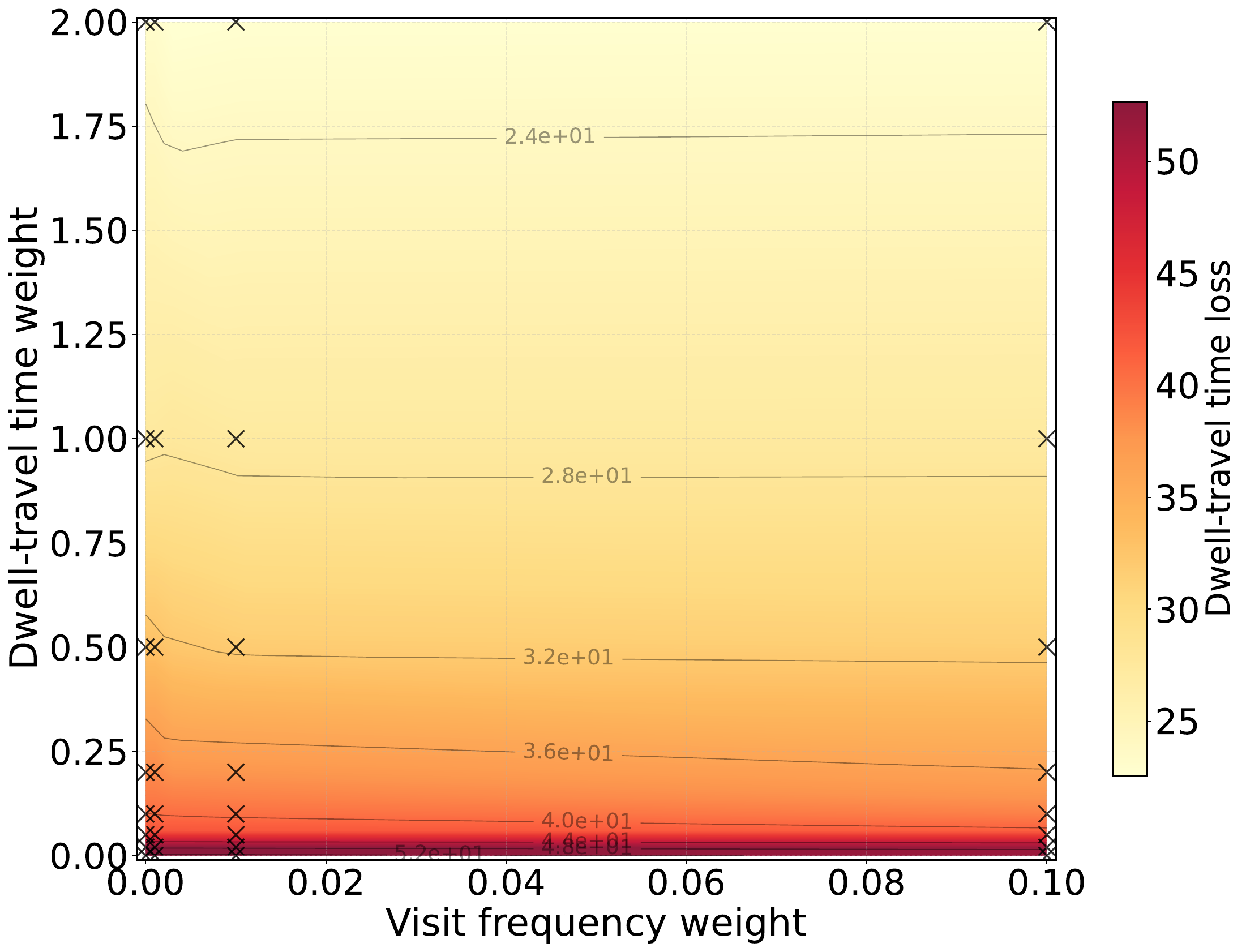}
\caption{\textbf{Loss functions $L^\mathrm{eval}_\mathrm{OD}$, $L_\mathrm{VF}$, and $L_\mathrm{DT}$ after optimization for Fukuoka Prefecture.}  
Each loss function was averaged over all attributes (sex and age groups). 
The horizontal and vertical axes represent $w_\mathrm{VF}$ and $w_\mathrm{DT}$, respectively. 
Black crosses indicate the simulated parameter combinations within the grid search range. All three plots demonstrated that the corresponding loss decreased as its associated weight increased.}
\label{fig:LossFunctionsFukuoka}
\end{figure*}

Figs.~\ref{fig:LossFunctionsTokyo} and \ref{fig:LossFunctionsFukuoka} present the values of the three loss functions, $L^\mathrm{eval}_\mathrm{OD}$, $L_\mathrm{VF}$, and $L_\mathrm{DT}$ for the Tokyo 23 special wards and Fukuoka Prefecture, respectively.  
All loss values were averaged over attributes (sex and age groups).  
The plots demonstrate that each loss function generally decreased as its corresponding weight increased, confirming that the optimization procedure functioned as intended.  
However, a trade-off was observed between the OD accuracy and dwell–travel time accuracy: the OD loss increased as the dwell–travel weight $w_\mathrm{DT}$ increased.  
In contrast, the visit frequency loss $L_\mathrm{VF}$ was largely independent of the other two metrics, suggesting that $w_\mathrm{VF}$ can be increased to improve the overall objective without adversely affecting the OD or dwell–travel time performance within the tested parameter range.

For the OD loss function, the square root of the modified loss, $\sqrt{L^\mathrm{eval}_\mathrm{OD}}$ represents the average relative error of the OD matrices.  
Figs.~\ref{fig:LossFunctionsTokyo}(a) and \ref{fig:LossFunctionsFukuoka}(a) indicate that this relative error changes, depending on parameter choices, from $\sqrt{0.0017} \approx 4.1\%$ to 
%$\sqrt{0.025} \approx 16\%$ 
$\sqrt{0.024} \approx 15\%$ 
in Tokyo and from 
%$\sqrt{0.0038} \approx 6.1\%$
$\sqrt{0.0037} \approx 6.1\%$
to $\sqrt{0.035} \approx 19\%$ in Fukuoka. 
The maximum acceptable relative errors of the OD matrices are discussed below.  
Shida \textit{et al.}~\cite{shida2022electric} reported that daily fluctuations in human mobility flows are on the order of their mean, corresponding to approximately $100\%$ relative error.  
According to the central limit theorem, the expected error in the OD matrices is therefore approximately $100\%/\sqrt{20} \approx 20\%$ given that the OD data were aggregated over approximately 20 weekdays.  
This value provides an upper bound for the permissible OD error in synthetic data.  
Although minimizing OD loss is desirable, overly small values would indicate overfitting of the OD data, leading to unbalanced synthetic mobility patterns.  
Thus, the optimization of the other criteria (visit frequency and dwell–travel distributions) within the acceptable error threshold was prioritized.  
In this study, the tolerance of the relative OD error was set to $10\%$, corresponding to $L^\mathrm{eval}_\mathrm{OD} < 0.01$.

The visit frequency loss function is defined as the Wasserstein distance between the synthetic and empirical visit frequency distributions averaged over all attributes, plotted in Fig. \ref{fig:LossFunctionsTokyo}(b) and \ref{fig:LossFunctionsFukuoka}(b). 
%The visit frequency loss function is defined as the Wasserstein distance between the synthetic and empirical visit frequency distributions averaged over all attributes.
This distance directly approximated the error in estimating visit frequencies from synthetic data.  
For the conventional GAD parameter settings $(w_\mathrm{OD}, w_\mathrm{VF}, w_\mathrm{DT}) = (1, 0, 0)$, the typical error was approximately $0.73$ in Tokyo and $0.65$ in Fukuoka. 
By selecting the optimal parameters for visit frequency reproduction, such as $(w_\mathrm{OD}, w_\mathrm{VF}, w_\mathrm{DT}) = (1, 0.1, 0.5)$, the error was reduced to $0.037$ for Tokyo and %$0.042$ for Fukuoka.  
$0.040$ for Fukuoka.  
In any case, the estimation error for the visit frequency can be successfully maintained below 
%$0.12$
$0.13$
because adjusting $w_\mathrm{VF}$ has a minimal impact on the other loss functions.

Finally, the dwell–travel loss function, defined as the weighted mean of the local Wasserstein distances for the dwell–travel time distributions, served as an estimator of the typical error in dwell–travel time computed from the synthetic data, plotted in Fig. \ref{fig:LossFunctionsTokyo}(c) and \ref{fig:LossFunctionsFukuoka}(c). 
%Finally, the dwell–travel loss function, defined as the weighted mean of the local Wasserstein distances for the dwell–travel time distributions, served as an estimator of the typical error in dwell–travel time computed from the synthetic data.  
In the conventional GAD parameter settings $(w_\mathrm{OD}, w_\mathrm{VF}, w_\mathrm{DT}) = (1, 0, 0)$, the average Wasserstein distance was $53$ min in both Tokyo and Fukuoka, which decreased to 
%$24$
$23$
min in both Tokyo and Fukuoka under the optimized parameter settings $(w_\mathrm{OD}, w_\mathrm{VF}, w_\mathrm{DT}) = (1, 0.1, 2)$.  
This indicates that the typical error was reduced by a factor of approximately 
%45\% 
43\% 
when the dwell–travel loss function was incorporated into the optimization.

In summary, the proposed method improved the accuracy of synthetic data for estimating visit frequency and dwell–travel time, while increasing the OD error only within the acceptable tolerance range.
The loss functions directly quantified the typical errors for each metric. 
For one of the best-performing parameter settings identified in the optimization, $(w_\mathrm{OD}, w_\mathrm{VF}, w_\mathrm{DT}) = (1, 0.1, 0.2)$, the OD error was $\leq 10\%$ in both the study areas.
In Tokyo, the visit frequency error was $0.058$ ($7.9\%$ of the original) and the dwell–travel time error was $33$ min ($63\%$ of the original), whereas in Fukuoka, the visit frequency error was $0.077$ ($12\%$ of the original) and the dwell–travel time error was $36$ min ($69\%$ of the original).

\section{Discussion and Conclusion}
This study presented a framework that generates synthetic mobility data when inputs are restricted to OD matrices and integrates auxiliary statistical information and universal laws as behavioral indicators. Specifically, temporal distributions were reconstructed using privacy-preserving quantile statistics of dwell–travel times and the daily visited location counts were aligned with the universal law reported in previous studies. By introducing both elements as auxiliary objectives within a multi-objective optimization procedure, this study demonstrated that the resulting data reproduced diverse behavioral properties under strict input constraints. For the first time, multiple behavioral constraints beyond OD consistency were simultaneously aligned within a framework directly applicable to industrial and governmental use cases. A modest decrease in the OD consistency was also observed and its magnitude was comparable to the day-to-day fluctuations intrinsic to human flow data, supporting the use of such fluctuations as a baseline for error assessment.

A distinctive feature of the proposed framework is its ability to safely incorporate external datasets using summary statistics. In this study, the quantile statistics of dwell–travel times newly provided by a parent company were used as auxiliary indicators. The same mechanism can be extended to datasets held by external organizations. For example, even records that are difficult to share in raw form, such as loyalty card purchase histories, can be aggregated into summary statistics and integrated into synthetic mobility data. This design allows stakeholders to combine their data with external sources in a privacy-preserving manner, expanding the range of potential applications, and increasing the value of synthetic data products for both enterprises and public agencies. Moreover, empirical regularities beyond human mobility, such as the scaling laws observed in purchasing behavior, can be incorporated similarly. By integrating external statistics and universal laws, synthetic mobility data can be tailored to specific analytical goals to achieve a higher practical utility. %Additional demographic or behavioral attributes can be incorporated as long as the input data are reliably stratified, with the usual sparsity–utility trade-off explicitly managed in the optimization. %%%Additional demographic,,,, camera ready

This study highlights the role of domain-wide ``stylized facts.” The proposed framework formalizes such laws as generalizable constraints, positioning them as building blocks for any future human mobility simulator, including large language model (LLM)-based and other generative models for complex time-series. In this view, even advanced synthetic generators should be regularized by fundamental empirical laws, and the proposed framework provides a principal mechanism for this integration.

Some limitations and opportunities for future work remain. First, there is an application-dependent trade-off between strict OD fidelity and alignment with auxiliary behavioral properties, which is affected by data noise, and the selection of appropriate weights is case-specific. Second, the framework depends on preprocessing choices, such as the 125 m grid resolution and treatment of short dwells, which may introduce bias under different sampling conditions. These aspects are not fundamental barriers but are areas for further refinement. Future work is planned to evaluate the robustness across additional regions and time periods and to assess the operational costs for large-scale deployments. It is also expected that the proposed design will inform institutional frameworks for data sharing, procurement, and regulatory compliance, thereby lowering the barriers for governments and enterprises to adopt synthetic mobility data responsibly at scale.

%%%%Although the raw GPS records were not accessible even for evaluation because of contractual restrictions, the framework was designed to remain externally verifiable. It relies solely on aggregated and summary-level statistics that can be compared with equivalent indicators computed from publicly available or third-party mobility datasets. This design ensures transparency and reproducibility without compromising data confidentiality.
%%%camera ready用に追加reviewer1-1 %%%Independent以下reviewr2のよー分からんコメントに
%%%%たりんので削除

Taken together, these results demonstrate that privacy-preserving synthetic mobility data can serve as a scalable and reliable foundation for urban analytics, supporting real-world deployments in both the policy and commercial domains without compromising individual privacy. %The framework is geography-agnostic and can be applied to any region where OD matrices and basic mobility statistics are available. %%%The framework is   camera ready%%イランから消した

Importantly, this study should be understood as a production-ready implementation rather than a conceptual proposal. Building on an existing commercial mobility data service, the proposed framework has already been implemented on production-scale datasets and directly enhances the value of the current data products of a company. As such, it is positioned for immediate deployment in governmental and industrial workflows, underscoring that the contribution goes beyond proof-of-concept and offers stakeholders a deployable solution while providing a foundation for future extensions.

\section*{Acknowledgements}
We would like to thank Editage (www.editage.jp) for English language editing.

\bibliographystyle{IEEEtran}
\bibliography{refs}        % refs.bib にエントリを書く

\end{document}